\documentclass[preprint,aps,amsmath,nofootinbib,tightenlines,floatfix]{revtex4}
\usepackage{graphicx,epsfig,amssymb}
\usepackage{multirow}
\usepackage{enumerate}

\begin{document}
\title{Trilepton Signals in the Inert Doublet Model}
\author{Xinyu Miao$^{1}$\footnote{miao@physics.arizona.edu},
  Shufang Su$^{1}$\footnote{shufang@physics.arizona.edu}, 
  Brooks Thomas$^{1,2}$\footnote{brooks@physics.arizona.edu}}
\affiliation{
 $^1$ Department of Physics, University of Arizona, Tucson, AZ  85721 USA\\
 $^2$ Department of Physics, University of Maryland, College Park, MD  20742 USA}

\begin{abstract}
In this work, we investigate the prospects for detecting the
Inert Doublet Model via the trilepton channel at the LHC.  We
present a set of representative benchmark scenarios in which
all applicable constraints are satisfied, and 
show that in some of these scenarios, it is possible to
obtain a signal at the $5\sigma$ significance level or better with
  integrated luminosity of $300\mbox{~fb}^{-1}$.
\end{abstract}

\maketitle

\newcommand{\newc}{\newcommand}
\newc{\gsim}{\lower.7ex\hbox{$\;\stackrel{\textstyle>}{\sim}\;$}}
\newc{\lsim}{\lower.7ex\hbox{$\;\stackrel{\textstyle<}{\sim}\;$}}

\def\beq{\begin{equation}}
\def\eeq{\end{equation}}
\def\beqn{\begin{eqnarray}}
\def\eeqn{\end{eqnarray}}
\def\calM{{\cal M}}
\def\calV{{\cal V}}
\def\calF{{\cal F}}
\def\half{{\textstyle{1\over 2}}}
\def\quarter{{\textstyle{1\over 4}}}
\def\ie{{\it i.e.}\/}
\def\eg{{\it e.g.}\/}
\def\etc{{\it etc}.\/}
\def\ET{{\displaystyle{\not}E_T}}


\def\inbar{\,\vrule height1.5ex width.4pt depth0pt}
\def\IR{\relax{\rm I\kern-.18em R}}
 \font\cmss=cmss10 \font\cmsss=cmss10 at 7pt
\def\IQ{\relax{\rm I\kern-.18em Q}}
\def\IZ{\relax\ifmmode\mathchoice
 {\hbox{\cmss Z\kern-.4em Z}}{\hbox{\cmss Z\kern-.4em Z}}
 {\lower.9pt\hbox{\cmsss Z\kern-.4em Z}}
 {\lower1.2pt\hbox{\cmsss Z\kern-.4em Z}}\else{\cmss Z\kern-.4em Z}\fi}




\section{Introduction\label{sec:Introduction}}

The Inert Doublet Model (IDM) is an extension of the Standard Model (SM) in which  
the scalar sector of the model includes, in addition to the Higgs field of the SM, 
an additional $SU(2)_L$ doublet (the so-called ``inert doublet") 
which couples only to the Higgs and electroweak gauge sectors of the SM.  
This additional doublet does not acquire a nonzero vacuum expectation value (VEV); 
hence the fields of this doublet do not contribute to electroweak-symmetry breaking, 
nor do they mix with those of the SM Higgs doublet. 
This model was originally proposed~\cite{FirstIDM} for its
applications to neutrino physics. 
However, the recent observation~\cite{BarbieriIDM} that the presence of an 
inert doublet can provide the necessary correction to the oblique $S$ and $T$
parameters~\cite{STU} to accommodate a heavy Higgs boson, with a mass as high as 
$400 -600$~GeV, without running afoul of LEP constraints~\cite{LEPBound} 
has stimulated a great deal of renewed interest in the model.
Since then, the IDM has found a host of additional applications to subjects as diverse 
as neutrino physics~\cite{IDMSeesaw}, electroweak-symmetry
breaking~\cite{BelgiansEWSBIDM}, and grand unification~\cite{IDMUnification}.
Furthermore, the model is also interesting from the point of view of dark-matter
phenomenology.  Since an additional discrete symmetry is needed in order to forbid the 
coupling of the inert doublet to SM fermions, the lightest physical degree of freedom 
in the inert doublet, if electrically neutral, is a viable Weakly Interacting Massive 
Particle (WIMP) dark-matter candidate.
Analyses of the relic abundance of this particle, often referred to as
the lightest inert particle (LIP), have been 
performed~\cite{BelgiansDarkIDM,ArizonansDarkIDM}, 
and assessments of both direct~\cite{DirectDetectIDM} and 
indirect (including neutrino~\cite{NeutrinoDarkIDM}, cosmic-ray
positron and antiproton~\cite{BelgiansPositronIDM},
and $\gamma$-ray~\cite{SwedesGammaRayIDM})
detection prospects have been carried out.  

Given the myriad implications and applications of this model, it is 
natural to examine how an inert doublet might be identified at the LHC.
Certain potential signatures were discussed in~\cite{BarbieriIDM}, 
and a parton-level analysis of some of these signatures (including a 
potentially observable modification of the total width of the SM 
Higgs boson) was performed in Ref.~\cite{CaoMa}.        
In Ref.~\cite{ArizonansDilepton}, a detailed, detector-level 
analysis of the detection prospects in the dilepton channel --- perhaps 
the most promising channel in which to search for evidence of an inert doublet 
at the LHC --- was performed for a set of representative benchmark points, each
corresponding to a particular region of model parameter space in which the LIP 
was able to account for the observed dark matter relic abundance.  This analysis 
showed that an observable signal in this channel could be obtained at the LHC with 
$100\mbox{~fb}^{-1}$ of integrated luminosity in a substantial number of these 
cases.      

Since a great many scenarios for physics beyond the Standard Model (BSM)
also give rise to a $\ell^+\ell^- + \displaystyle{\not}E_T$ signature, 
it is worthwhile to investigate other channels which might also yield 
observable signals indicating the presence of an inert doublet.     
In this work, we focus on the detection prospects in the trilepton channel: 
$\ell^+\ell^-\ell^\pm + \displaystyle{\not}E_T$.
Indeed, this channel has long been regarded as one of the most
promising channels in which to look for evidence of physics beyond the
Standard Model, and, in particular, of supersymmetry~\cite{trilepton},
due to its relatively small SM background.
   
We begin in Sect.~\ref{sec:Model} with a brief review of the Inert Doublet 
Model and a summary of the relevant theoretical and experimental constraints.  We  
then present a set of benchmark scenarios in which all of these constraints are
satisfied and the LIP relic density accounts for the dark matter abundance 
observed by WMAP~\cite{WMAP2003}.  
In Sect.~\ref{sec:Cuts}, we discuss trilepton production in the IDM and
outline the event-selection criteria we use to differentiate the 
trilepton signal from the SM background.  In Sect.~\ref{sec:Results}, we
present our numerical results and discuss the LHC discovery potential for 
each of our benchmark scenarios, and in 
Sect.~\ref{sec:Conclusion}, we offer some concluding remarks on the scope
and implications of our results. 
  
\section{Model Parameters and Benchmark Points\label{sec:Model}}

The Inert Doublet Model is an extension of 
the SM in which the fundamental scalar sector comprises not one,
but two scalar doublets, which we denote $\phi_1$ and $\phi_2$. 
The first of these doublets, $\phi_1$, can be identified with the usual 
Higgs doublet of the SM.  It receives a nonzero 
VEV of $\langle\phi_1^0\rangle = v/\sqrt{2} =
174$~GeV, and consequently bears full responsibility both for electroweak-symmetry 
breaking (EWSB) and for the generation of SM fermion masses.  By contrast, the second 
doublet, $\phi_2$, does not contribute to electroweak-symmetry breaking
($\langle \phi^0_2\rangle = 0$); furthermore, it is prohibited from coupling to the 
quarks and leptons of the SM by a $\mathbb{Z}_2$ parity (often called matter parity) 
under which $\phi_2$ is odd,
whereas all other fields in the theory (including $\phi_1$) are even.  This 
$\mathbb{Z}_2$ symmetry also renders the lightest physical degree of freedom
contained in $\phi_2$ --- the so-called lightest inert particle, or LIP --- 
absolutely stable, and hence (if neutral) a good WIMP dark-matter candidate.

The most general $CP$-invariant scalar potential which respects both 
$SU(2)_L\times U(1)_Y$ gauge-invariance and the aforementioned $\mathbb{Z}_2$ 
matter parity may be written in the form 
\begin{equation}
   V = \mu_1^2|\phi_1|^2 + \mu_2^2|\phi_2|^2 + \lambda_1|\phi_1|^4 + \lambda_2|\phi_2|^4
      + \lambda_3|\phi_1|^2|\phi_2|^2 + \lambda_4 |\phi_1^\dagger\phi_2|^2 
      + \left[\frac{\lambda_5}{2}(\phi_1^\dagger\phi_2)^2 +\mathrm{h.c.}\right].
   \label{eq:VScal}
\end{equation}   
After EWSB is triggered by the VEV of $\phi_1$, the physical scalar spectrum of the model comprises
the usual SM Higgs field $h$ (the neutral, $CP$-even degree of freedom in $\phi_1$), as well 
as four additional fields corresponding to the four degrees of freedom in $\phi_2$.  These
include a pair of charged scalars $H^\pm$, a neutral, $CP$-even scalar $S$, and a neutral,
$CP$-odd scalar $A$.  The masses of these scalars, written in terms of the parameters
appearing in Eq.~(\ref{eq:VScal}) and the SM Higgs VEV $v$, are given by
\begin{eqnarray}
m_{h}^2 &=& 2\lambda_1 v^2,\\
m_{H^\pm}^2 &=& \mu_2^2 + \lambda_3 v^2/2,\\
m_S^2 &=& \mu_2^2 + (\lambda_3 +\lambda_4 +\lambda_5 ) v^2/2,\\
m_A^2 &=& \mu_2^2 + (\lambda_3 +\lambda_4 -\lambda_5 ) v^2/2.
\end{eqnarray}     
It is also  useful to define a pair of mass splittings $\delta_1\equiv m_{H^\pm} - m_S$
and $\delta_2\equiv m_{A} - m_S$.  Indeed, for our present purposes, it is most 
convenient to characterize a given model in terms of the parameter set 
$\{m_h,m_S,\delta_1,\delta_2,\lambda_2,\lambda_L\}$, where
$\lambda_L \equiv \lambda_3 + \lambda_4 +\lambda_5$ represents the combination of 
the $\lambda_i$ in Eq.~(\ref{eq:VScal}) which controls the size of the $hSS$ and $hhSS$ 
couplings.

The parameter space of the IDM is restricted by a number of considerations.  These include 
model-consistency requirements, such as perturbativity and vacuum stability, as well as 
experimental constraints derived from electroweak precision data, the results of direct 
searches at LEP and at the Tevatron, etc.  In addition, if we further demand 
that the LIP be the dominant component of the dark matter, we must also require that its
relic density falls within the WMAP $3\sigma$ bounds~\cite{WMAP2003} on the observed 
dark-matter relic density, and that the appropriate constraints from dark-matter 
direct-detection experiments are likewise satisfied.  
Since these constraints, along with their implications for the parameter space of the IDM, 
have been analyzed and discussed in detail in Refs.~\cite{ArizonansDarkIDM,ArizonansDilepton}, 
we will not recapitulate the analysis here, but simply summarize the results.
Note that in our analysis below, we will assume that the $CP$-even scalar $S$ is the dark matter candidate.  The corresponding results obtained in the case in which $A$ plays the role of the 
dark matter candidate are very similar.

\begin{table}
\begin{center}
\begin{tabular}{|c|c|c|c|c|c|}\hline
~~Benchmark~~ & ~$m_h$ (GeV)~& ~$m_S$ (GeV)~ & ~$\delta_1$ (GeV)~ & 
~$\delta_2$ (GeV)~ & ~~~~$\lambda_L$~~~~  \\\hline
LH1& 150 &  40 & 100 & 100 & $-0.275$\\
LH2& 120 &  40 &  70 &  70 & $-0.15$ \\
LH3& 120 &  82 &  50 &  50 & $-0.20$ \\
LH6& 130 &  40 & 100 &  70 & $-0.18$ \\
LH7& 117 &  37 &  70 & 100 & $-0.14$ \\
LH8& 120 &  78 &  70 &  35 & $-0.18$ \\\hline
\end{tabular}
\caption{A list of benchmark points used in our analysis, defined in terms of the model parameters $\{m_h,m_S,\delta_1,\delta_2,\lambda_L\}$.  Dark matter relic density and collider phenomenology of the IDM depend little on $\lambda_2$, which is set to 0.1  for all benchmark points in this study.}
\label{tab:BMs}
\end{center}
\end{table}

A number of IDM parameter-space regimes exist in which all of the aforementioned physical
constraints are satisfied, and in which the LIP can account for the observed dark-matter
abundance. 
In the first of these regimes, dubbed the ``LH'' or ``light-Higgs'' regime, 
the SM Higgs mass lies within the 
$114\mbox{~GeV}\lesssim m_h \lesssim 186~\mbox{GeV}$ preferred by LEP constraints, and 
electroweak precision data require
that $\delta_1$ and $\delta_2$ be of roughly the same order.  Here, the LIP is light
enough, with a mass in the range $35\mbox{~GeV}\lesssim m_S \lesssim 80\mbox{~GeV}$,
that its relic abundance is not washed out by $SS\rightarrow WW^{(\ast)}$ annihilation.

A number of benchmark points
corresponding to different possible scenarios within this regime were defined in Ref.~\cite{ArizonansDilepton}.  These were specifically
chosen with the dilepton channel in mind.  
The parameter assignments for some of those points, labeled
LH1 $-$ LH3, are listed in Table~\ref{tab:BMs}.  LH1 and LH2 represent scenarios in which 
the mass of the dark-matter candidate is light ($m_S \sim$ 40~GeV), and in which the
relationships between $M_W$, $M_Z$, and the mass splittings $\delta_1$ and $\delta_2$ are 
given by $\delta_1=\delta_2>M_{W,Z}$ and $\delta_1=\delta_2<M_{W,Z}$ for point LH1 and LH2,
respectively.  
Since the contributions from the SM background (and consequently the event-selection 
criteria we impose) differ significantly depending on whether the $W$ and $Z$ bosons 
during the decays of the inert scalars are on- or off-shell, 
we also incorporate two more benchmark points into our analysis.  These points, which we
dub LH6 and LH7, respectively 
represent the situations in which $\delta_1 >M_{W}$, with $\delta_2 <M_{Z}$; and $\delta_1 <M_{W}$, with $\delta_2 >M_{Z}$.
Benchmarks LH3 and LH8 represent scenarios in which the mass of the dark-matter candidate 
is roughly $m_S\sim 80$ GeV, and in which both $\delta_1$ and $\delta_2$ are restricted 
to be lighter than $M_{W}$.  LH3 represents the case in which $\delta_1 \sim \delta_2$, 
while LH8 represents the case when $\delta_1 > \delta_2$.  The latter case has been included
primarily as it exemplifies the case in which $SH^\pm$ pair production, with 
$H^\pm \rightarrow A W^{\pm *} \rightarrow S Z^* W^{\pm *}$, also contributes 
to the trilepton signal.   Note that point LH6 $-$ LH8 have been included 
(in addition to the benchmark points defined in Ref.~\cite{ArizonansDilepton}) to 
highlight the effect of relationships between model parameters (primarily 
$\delta_1$ and $\delta_2$) which have little effect on results in the 
dilepton channel (the results in which are fairly insensitive to the value 
of $\delta_1$), but have a substantial effect on results in the trilepton channel.    
The additional benchmark points LH4 and LH5 defined in Ref.~\cite{ArizonansDilepton} have
been left out of this analysis, since each of those two points involves a value of 
either $\delta_1$ or $\delta_2$ so small (10~GeV) that the vast majority of final-state 
leptons will escape detection, rendering a trilepton signal essentially unobservable at
the LHC. 

In the second parameter-space regime in which the IDM successfully accounts for the
observed dark-matter abundance, which we dub 
the ``HH'' or ``heavy-Higgs'' regime, the Higgs mass lies within
the range $400\mbox{~GeV}\lesssim m_h\lesssim 600\mbox{~GeV}$.  In this case,
electroweak-precision constraints require that $\delta_1 \gg \delta_2$, with 
$\delta_1$ typically larger than $150$~GeV.   Consequently, the charged
scalars $H^\pm$ tend to be quite heavy.  In addition, in order to satisfy the WMAP
constraint on the dark-matter density, the LIP mass must be $m_S \approx 75$~GeV.  While
the prospects for detecting a dilepton signature in such scenarios can be quite 
good~\cite{ArizonansDilepton}, depending primarily on whether or not $\delta_2 < M_Z$,
a trilepton signature proves far more difficult to detect.   
The reason for this is that the primary source of observable dilepton events in the 
IDM is the pair-production process $pp\rightarrow SA$, the cross-section for which is
independent of $\delta_1$.  By contrast, the primary source for an observable trilepton 
signal, as will be discussed in more detail below, is $pp\rightarrow H^\pm A$, which depends
on both $\delta_1$ and $\delta_2$.  The cross-section for this process is therefore
substantially suppressed due to the large value of $\delta_1$.  Consequently, in what 
follows, we will focus primarily on the detection 
prospects for the light-Higgs scenarios alone.         
 
In addition to these two cases, a number of other parameter-space regimes exist in which
all constraints are satisfied, and in which the LIP relic density reproduces the observed 
dark matter relic density~\cite{ArizonansDarkIDM}.  However, these scenarios do not yield
observable signals in either the dilepton or trilepton channel, generally either because
all of the inert scalars are required to be extremely heavy, or else because $\delta_2$ is   
required to be quite small ($\delta_2 \lesssim 30$~GeV), and hence the charged leptons
resulting from $A$ decay are extremely soft.  We will therefore not consider the collider phenomenology of such models
further here, but we emphasize that they are still viable scenarios.

\section{Trilepton Production in the Inert Doublet Model\label{sec:Cuts}}

\begin{figure}[bht]
\begin{center}
\resizebox{2.5in}{!}{\includegraphics{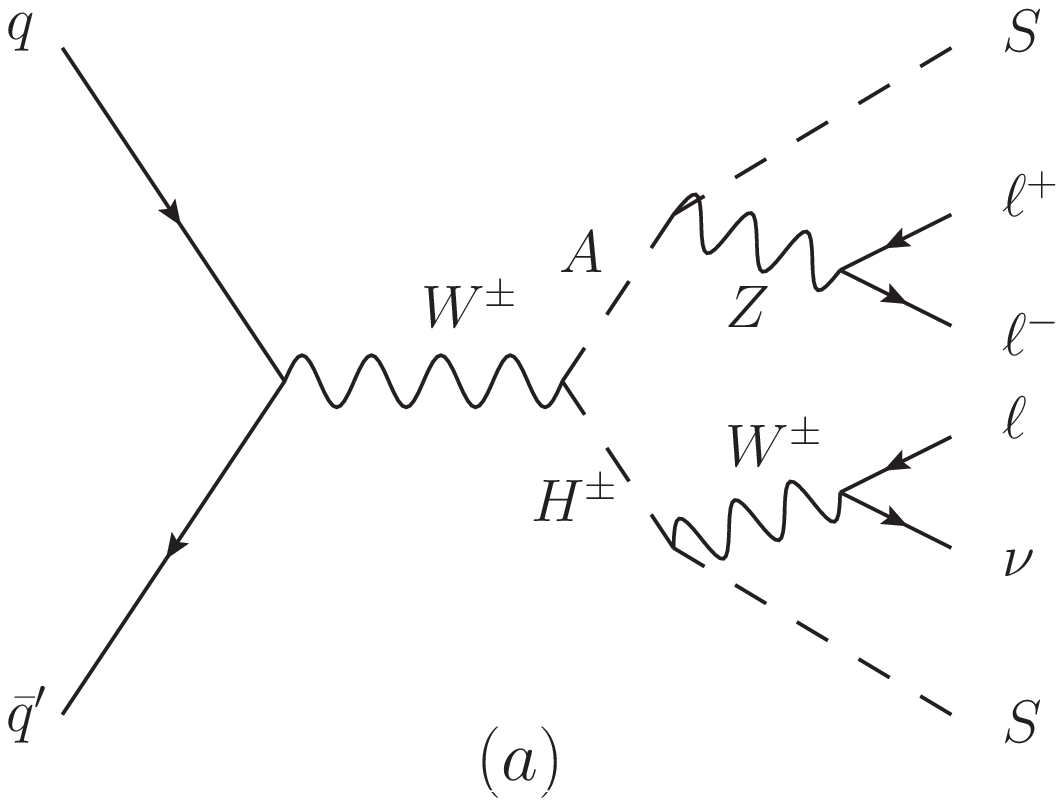}}~~~~~~~~~
\resizebox{2.5in}{!}{\includegraphics{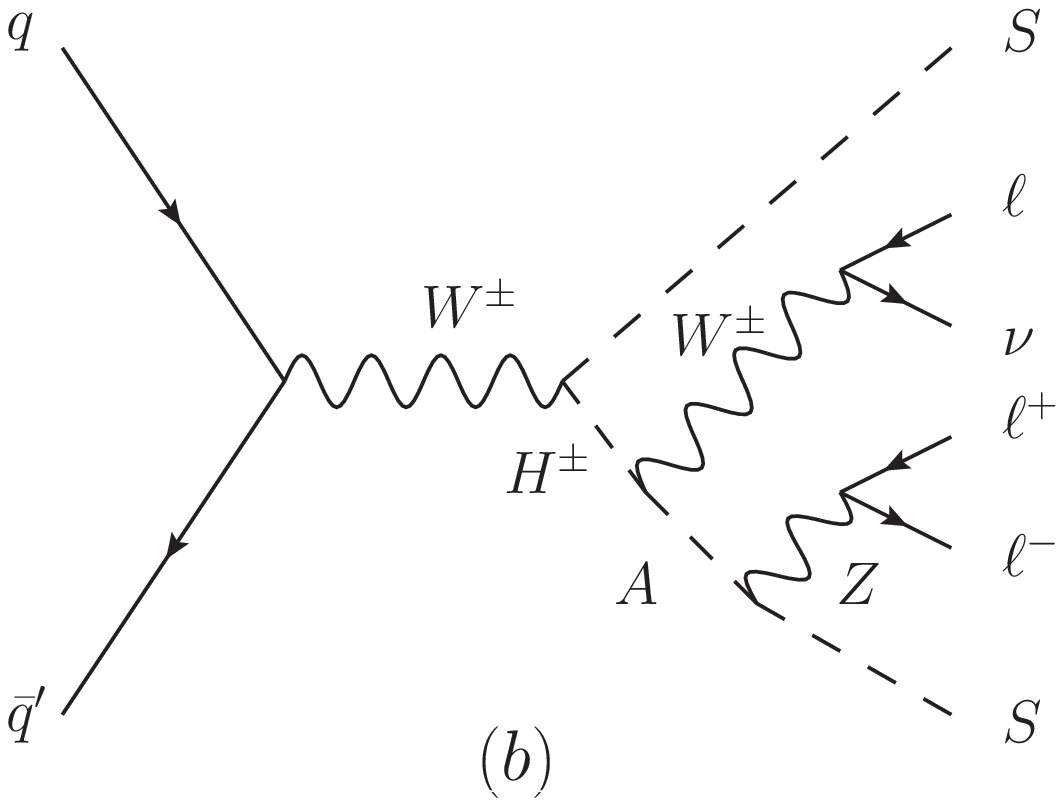}}
\caption{Diagrams corresponding to the processes which provide the leading
contributions to the
$\ell^+\ell^-\ell^\pm +\displaystyle{\not}E_T$ cross-section in the IDM.
\label{fig:TrilepDiagrams}}
\end{center}
\end{figure}

A number of processes contribute to the overall trilepton signal in the IDM.  Here,
we will concentrate on the most promising contributions for detection: those in which
one lepton is produced via $W^{(*)}$ decay and the other two via $Z^{(*)}$ decay.
The most significant such contributions are
\begin{enumerate}[(a)]
\item{} $q\bar{q}^\prime \rightarrow A H^\pm$ with $A\rightarrow S Z^{(*)} \rightarrow S \ell^+ \ell^-$ and $H^\pm\rightarrow S W^{\pm(*)} \rightarrow S \ell \nu$;
\item{} $q\bar{q}^\prime \rightarrow S H^\pm$ with  $H^\pm\rightarrow A W^{\pm(*)} \rightarrow A \ell \nu$ and  $A\rightarrow S Z^{(*)} \rightarrow S \ell^+ \ell^-$,
\end{enumerate}
the corresponding Feynman Diagrams for which are shown in Fig.~\ref{fig:TrilepDiagrams}.
Note that in our analysis, we will consider the case in which $\ell=e,\mu$ only.

Process (a) will occur in any IDM scenario in which the $S$ plays the role
of the LIP,  whereas process (b) will occur only in scenarios in which 
$\delta_1 > \delta_2$ and will only be sizeable when 
$\delta_1 < M_W$ or $\delta_1 > \delta_2+M_W$.
For all the benchmark points listed in Table~\ref{tab:BMs}, process (b) is sizable only 
for LH8.  Even in that case, it is subdominant compared to process (a), the overall
cross-section for which (taking into account all relevant decay branching ratios) 
is a factor of 20 larger than that for process (b).
In cases in which $\delta_2 > \delta_1$, the process $q\bar{q}^\prime \rightarrow A H^\pm$, 
with $A\rightarrow H^\pm W^{\mp(*)} \rightarrow H^\pm \ell \nu$ and 
$H^\pm\rightarrow S W^{\pm(*)} \rightarrow S \ell \nu$, also contributes to trilepton 
production.  The leptons produced in this process all come from $W^{(*)}$ decay, and
for this reason, it is difficult to resolve this process from the SM background.
For all the benchmark points that we have selected for our study, however, the overall 
cross-section for this process is negligibly small, and can therefore be safely neglected.

Results for the LHC production cross-sections
for the dominant ($pp\rightarrow AH^\pm$) signal process at $\sqrt{s}=14$ TeV, 
as well as the branching fractions for $H^\pm\rightarrow S\ell^\pm\nu$ 
and $A\rightarrow S\ell^+\ell^-$ decay, are provided in Table~\ref{tab:BMXSecs}.  
Note that for benchmark point LH8, the subdominant contribution to the trilepton 
signal from $pp\rightarrow SA^\pm$ has also been included in our analysis.
 
\begin{table}
\begin{center}
\begin{tabular}{|c|c|cc|}\hline
~~Benchmark~~ &~$\sigma_{AH^\pm}$  (fb) ~
& ~ $\mathrm{BR}(H^\pm\rightarrow S\ell^\pm\nu)$ ~ 
& ~ $\mathrm{BR}(A\rightarrow S\ell^+\ell^-)$ ~ \\
 \hline
LH1 &   125.2     & 0.216 & 0.067 \\
LH2  & 299.0     & 0.233 & 0.068 \\
LH3  & 154.9     & 0.233 & 0.069 \\
LH6  & 187.0    & 0.216 & 0.069 \\
LH7  & 204.2   & 0.233 & 0.067 \\
LH8  & 159.4  & 0.226 & 0.070 \\\hline
\end{tabular}
\caption{Leading-order cross-sections for the associated production of 
 $AH^\pm$ at the LHC, with center-of-mass energy 
$\sqrt{s}=14$~TeV, for the various benchmark points defined in 
Table~\ref{tab:BMs}.  The relevant branching fractions of the
scalars $A$ and $H^\pm$ are also shown.  
\label{tab:BMXSecs}}
\end{center}
\end{table}

A number of processes contribute to the SM background for trilepton production.
The most important of these is the irreducible background from $WZ/\gamma^\ast$
production, though a number of reducible backgrounds also contribute.  These
include $t\bar{t}(j)$, $Wt(j)$, $ZZ$, and, as recently emphasized 
in~\cite{trilepheavyflavor}, heavy-flavor processes such as
$b\bar{b}Z/\gamma^*$ and $c\bar{c}Z/\gamma^*$.  

In our analysis, event samples both for the signal process and for these backgrounds 
were generated at parton-level using the MadGraph~\cite{MadGraph}
package.  These events were subsequently passed through PYTHIA~\cite{PYTHIA} 
for parton showering and hadronization, and then through PGS4~\cite{pgs} to 
simulate the effects of a realistic detector.  The one exception involves 
the background from heavy-flavor processes, which is somewhat cumbersome
to analyze numerically, given the amount of data required to obtain a
statistically reliable sample.   
However, as has been shown in Ref.~\cite{trilepheavyflavor}, this background
can be effectively eliminated via the implementation of a stringent
missing energy cut of order $\displaystyle{\not}E_T> 50$~GeV.
A similarly stringent cut on the total transverse 
momentum variable $H_T$  
should also be quite effective in this regard.  We shall therefore assume 
that these backgrounds are effectively eliminated by the 
$\displaystyle{\not}E_T$ and $H_T$ cuts included among our event-selection
criteria.

Let us now turn to discuss those event-selection criteria, which we apply in
three successive stages or sets, in more detail.
The first set of cuts we impose (hereafter to be referred to as our Level I cuts) 
is designed to mimic a realistic detector acceptance.  More specifically, 
we require:
\begin{itemize}
\item Exactly three charged leptons (either electrons or muons), including
   one same-flavor, opposite-sign (SFOS) pair.
\item $p_T^\ell>15$~GeV and $|\eta_\ell|<2.5$ for each of these leptons.
\item For lepton isolation, we require $\Delta R_{\ell\ell} > 0.4$ for each possible 
   charged-lepton pairing, and $\Delta R_{j\ell}>0.4$ for each combination of one 
   jet and one charged lepton.
\end{itemize}    

Our second set of cuts (hereafter referred to as our Level II cuts) 
is designed to suppress reducible
backgrounds from SM processes which involve either hard jets or little
missing transverse energy:  
\begin{itemize}
\item No jets with $p_T^j > 20$~GeV and pseudorapidity $|\eta|<3.0.$ 
\item $\ET > 50$~GeV.
\end{itemize}  
As discussed above, a missing-energy cut of this magnitude 
effectively eliminates the background from heavy-flavor processes 
such as $b\bar{b}Z/\gamma^\ast$ and $c\bar{c}Z/\gamma^\ast$.  The jet veto is
quite efficient in reducing background contributions from
$t\bar{t}(j)$, $Wt(j)$, and other processes which involve substantial
hadronic activity in the central region of the detector.
Indeed, after the application of the Level~I+II cuts discussed above, the dominant
remaining background is the irreducible one from $WZ/\gamma^\ast$ production, as
shown in Table~\ref{tab:LevelIICuts}.
In addition, there is also a non-negligible contribution (amounting to around 5\% 
of the $WZ/\gamma^*$ background) from residual $t\bar{t}(j)$ and $Wt(j)$ events 
which survive the jet veto.
Other reducible backgrounds, including those from $W$ + jets and 
heavy-flavor processes, are effectively eliminated by
this choice of cuts.

\begin{table}
\begin{center}
\begin{tabular}{|c|cc|c|cc|}\hline
 \multicolumn{3}{|c|}{Signal} & \multicolumn{3}{c|}{SM Background} \\ \hline   
\multirow{2}{*}{~Benchmark~} & ~Level~I~ &  ~Level~I+II~~ &
\multirow{2}{*}{~Process~}   & ~Level~I~ &  ~Level~I+II~~ \\
&(fb)&(fb)&&(fb)&(fb) \\ \hline
LH1 & 0.760	& 0.317 & $W Z/\gamma^\ast$  & 125.767 & 32.949 \\
LH2 & 0.817 & 0.290 & $t\bar{t}(j)$         & 38.869 &  1.046 \\
LH3 & 0.289 & 0.082 & $Wt(j)$              & 1.794 &  0.536 \\
LH6 & 0.618 & 0.239 & Total BG           & 166.430 & 34.531 \\  
LH7 & 1.089	& 0.420 &  &  & \\
LH8 & 0.204 & 0.048 &  &  &   \\ \hline
\end{tabular}
\caption{Cross-sections for the signal process
$pp\rightarrow AH^\pm\rightarrow \ell^+\ell^-\ell^\pm + \displaystyle{\not} E_T$
in each of the benchmark scenarios presented in Table~\ref{tab:BMs}, and for the
relevant SM backgrounds, after the application of our Level~I and Level~II cuts.
\label{tab:LevelIICuts}}
\end{center}
\end{table}

After the imposition of the Level~I and Level~II cuts, we impose one further battery
of event-selection criteria (hereafter referred to as our Level III cuts).  
Unlike these first two sets of cuts, which are applied 
universally to all benchmark points used in this analysis, our Level~III cuts are 
individually tailored to optimize the statistical significance of discovery for 
each benchmark point.  A wide variety of possible criteria could in principle be used
in this optimization process; however, we find that 
one particularly useful criterion that can be used to differentiate between signal
and background events is the invariant mass $M_{\ell_Z\ell_Z}$ of the requisite pair 
of SFOS charged leptons (which we dub $\ell^+_Z$ and $\ell^-_Z$) that any event must 
include in order to pass the Level~I cuts.  If only one SFOS pairing can be constructed 
for a given event, $M_{\ell_Z\ell_Z}$ is unambiguously defined.
In cases in which more than one SFOS combination exists and $\delta_2\geq 70$~GeV, 
the pair whose invariant mass is closest to $\min(\delta_2,M_Z)$ will be 
identified as $\ell^+_Z$ and $\ell^-_Z$, and that 
invariant mass will be identified as $M_{\ell_Z\ell_Z}$.  In cases in which 
$\delta_2 < 70$~GeV, the pair whose invariant mass is closest to $70$~GeV will
be so identified.\footnote{We choose this criterion for identifying the SFOS pair, 
rather that simply selecting whichever pair has an invariant mass closer to $\delta_2$.
This is
because for $\delta_2\lesssim 70$~GeV, the latter procedure would result in more frequent
misidentification of which leptons were produced via $Z/\gamma^\ast$ decay in the
$WZ/\gamma^\ast$ background sample, and consequently lower statistical significance values.}

\begin{figure}[bht]
\begin{center}
\resizebox{3.15in}{!}{\includegraphics{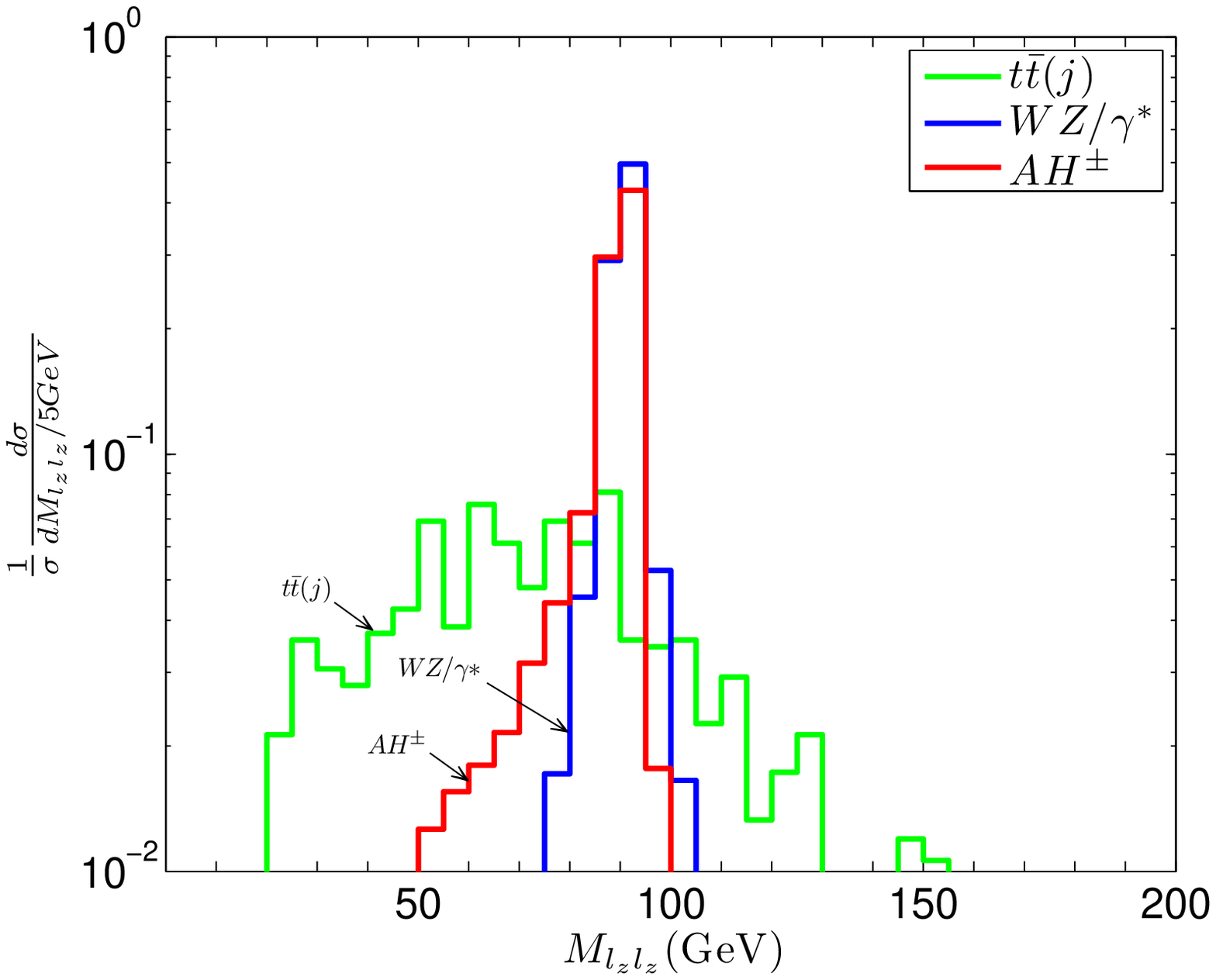}}
\resizebox{3.15in}{!}{\includegraphics{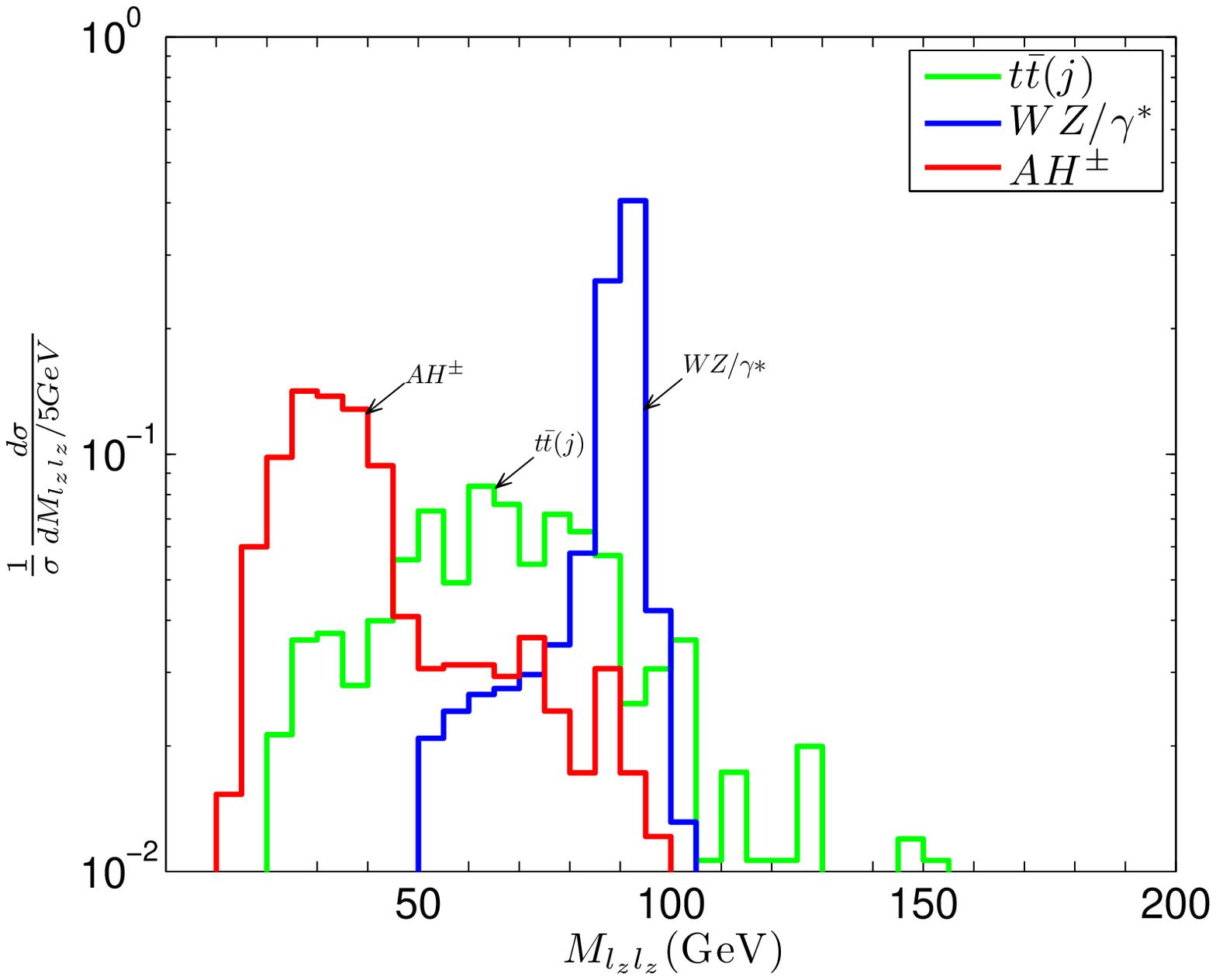}}
\caption{Distributions of the invariant mass of the SFOS 
lepton pair after the application of the Level~I+II cuts described in the
text, in our benchmark scenarios LH1 (left panel) and LH3 (right panel), 
both for the signal process and for the dominant SM backgrounds.  Note that
the area under each distribution curve has been normalized to one.        
\label{fig:mll}}
\end{center}
\end{figure}

The distribution for $M_{\ell_Z\ell_Z}$ peaks around $M_Z$ for the Standard-Model 
$WZ/\gamma^\ast$ background.  
For the signal process, the peak is around ${\rm min}(\delta_2, M_Z)$, 
as shown clearly in Fig.~\ref{fig:mll} for LH1 ($\delta_2=100$ GeV, left panel) and 
LH3 ($\delta_2=50$ GeV, right panel).  This suggests that a cut on 
$M_{\ell_Z\ell_Z}$ around $\delta_2$ has the potential to suppress 
significantly the SM background in scenarios in which $\delta_2<M_Z$.  
Therefore, in our analysis, we select events on the basis of whether 
$M_{\ell_Z\ell_Z}$ lies below the threshold      
\begin{itemize}
\item $M_{\ell_Z\ell_Z} 
          \leq M_{\ell_Z\ell_Z}^{\mathrm{max}}$.
\end{itemize}
In principle, one could also introduce a minimum threshold for $M_{\ell_Z\ell_Z}$, but it turns
out that the imposition of such a cut is not particularly helpful in practice; thus we 
will only make use of the above criterion in what follows.

Furthermore, in cases in which $A\rightarrow S\ell^+\ell^-$ decay occurs 
via an off-shell $Z$, the charged leptons will tend to be more collinear than those produced
from the decay of an on-shell $Z$.  For this reason, cuts such as
\begin{itemize}
\item $\cos\phi_{\ell\ell}\geq\cos\phi_{\ell\ell}^{\mathrm{min}}$  
\item $\Delta R_{\ell\ell}\leq\Delta R_{\ell\ell}^{\mathrm{max}}$, 
\end{itemize}   
where $\phi_{\ell\ell}$ is the azimuthal angle between the SFOS lepton pair, can be
quite effective in discriminating between signal and background in cases in which
$\delta_2<M_Z$.  In practice, we find the $\Delta R_{\ell\ell}^{\mathrm{max}}$
cut alone to be sufficient for our purposes, and thus make use of this criterion
exclusively. 

\begin{figure}[bht]
\begin{center}
\resizebox{3.15in}{!}{\includegraphics{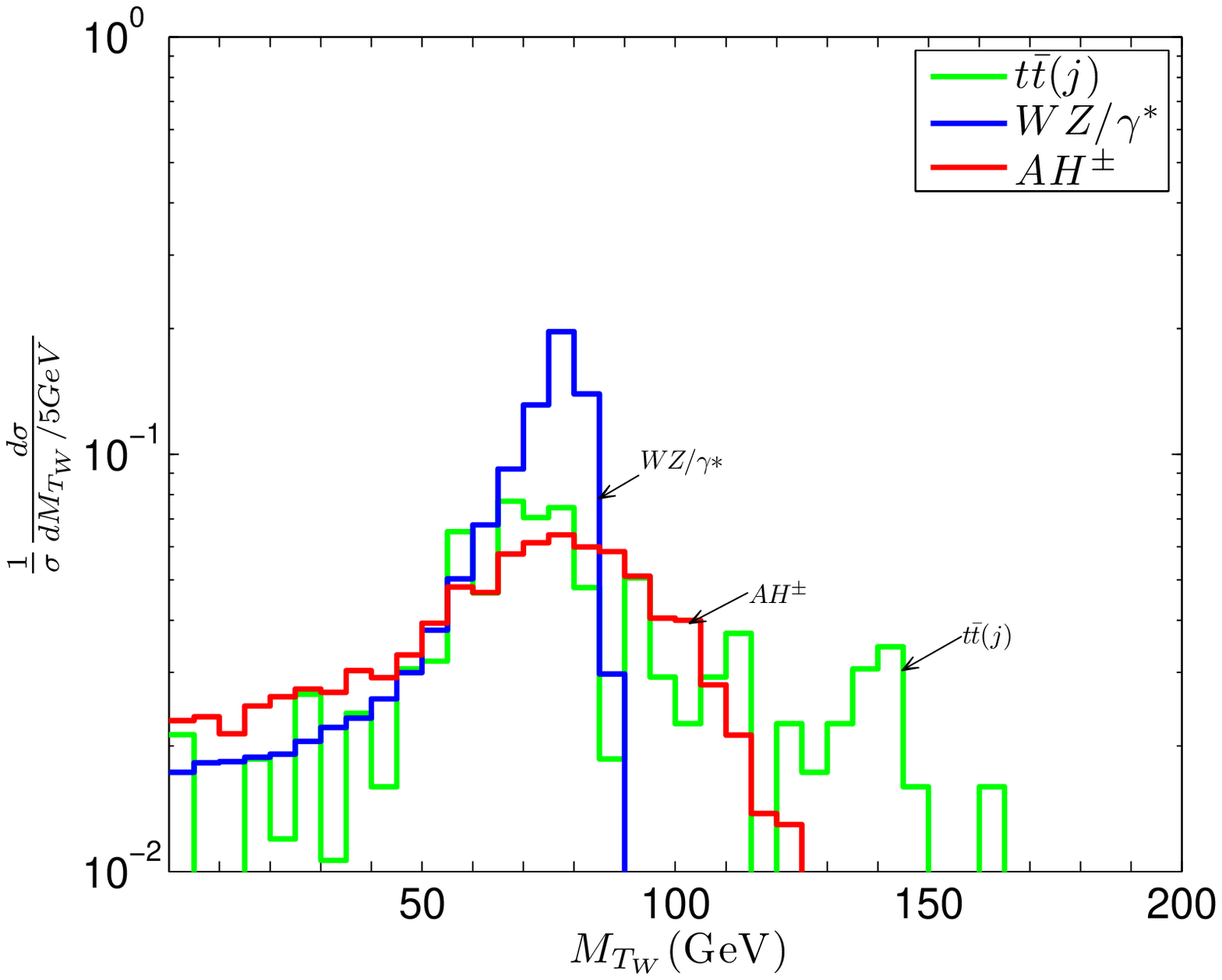}}
\resizebox{3.15in}{!}{\includegraphics{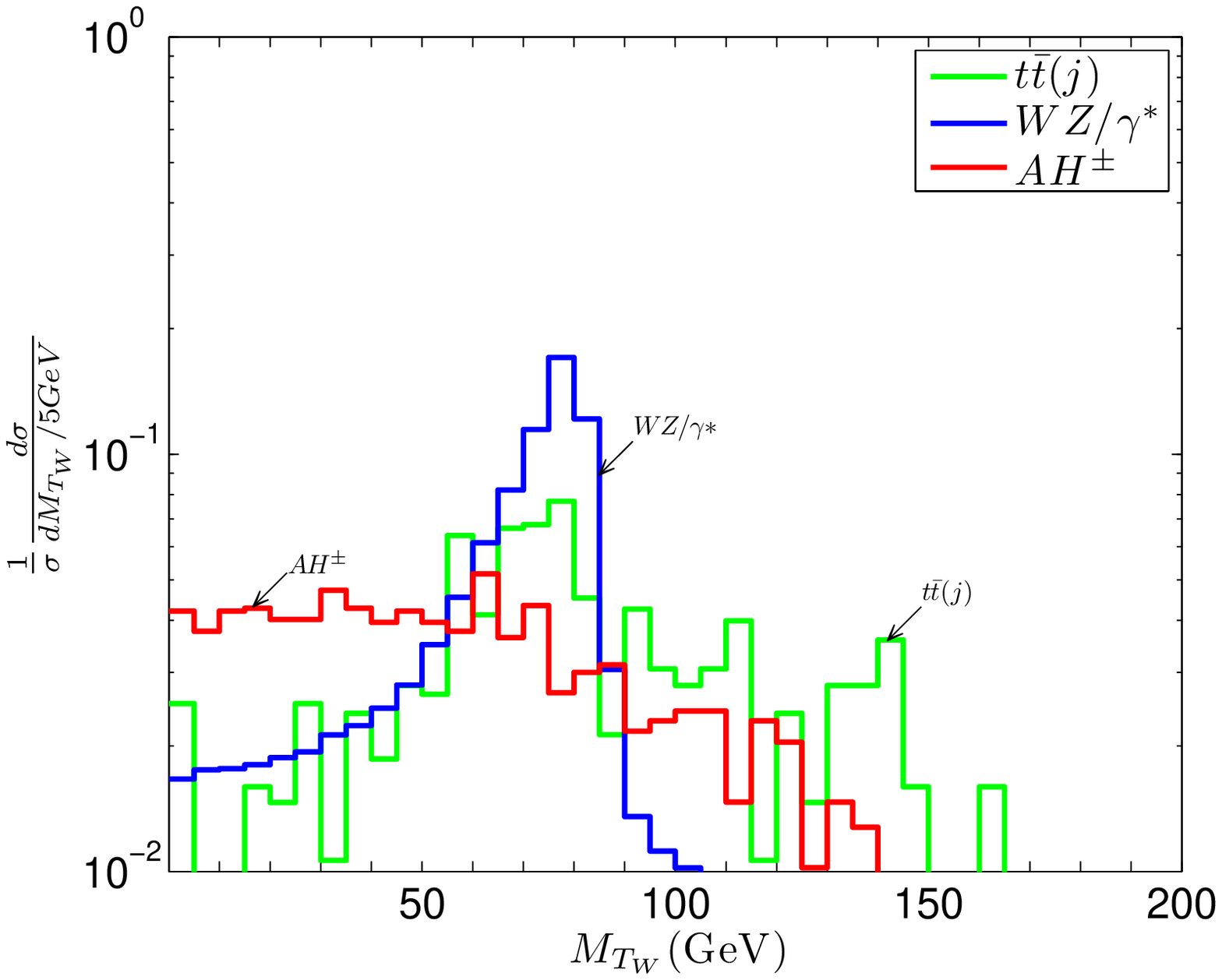}}
\caption{Distributions of the transverse mass variable $M_{T_W}$ defined
in Eq.~(\ref{eq:MtwDef}), after the application of the Level~I+II cuts 
described in the text, in our benchmark scenarios LH1 (left panel) 
and LH3 (right panel), both for the signal process and for the dominant 
SM backgrounds.       
\label{fig:mTW}}
\end{center}
\end{figure}

From the four-momentum of the remaining lepton (the one that is not part
of the $\ell_Z^+\ell_Z^-$ pair), which we dub $\ell_W$, 
we can construct an additional quantity: a transverse-mass variable $M_{T_W}$, 
which we define according to the relation   
\begin{equation}
  M_{T_W}^2 \equiv (E_{\ell_W} + \displaystyle{\not}E_T)^2 - 
             (\vec{p}_{T\ell_W} + \displaystyle{\not}\vec{p}_T)^2,
  \label{eq:MtwDef}
\end{equation}
where $\displaystyle{\not}E_T$ and $\displaystyle{\not}\vec{p}_T$ respectively denote the
{\it total} missing transverse energy and missing transverse momentum vector.  
The distribution for $M_{T_W}$ drops sharply around $M_W$ for the SM $WZ/\gamma^*$ background.
A similar drop also occurs for the signal process, in cases in which the $H^\pm$ decays via 
an on-shell $W$,
but the presence of additional sources of $\displaystyle{\not}E_T$ 
(the pair of LIPs) in this case results in a smoother $M_{T_W}$ distribution that 
falls more gently above $M_W$.  In cases in which $\delta_1<M_W$, and the lepton in
question comes from off-shell $W$ decay, the drop in $M_{T_W}$ is quite gradual and 
occurs near $\delta_1$.
The distributions for $M_{T_W}$, both for the signal process and for the dominant SM
backgrounds, are shown in Fig.~\ref{fig:mTW} for LH1 
($\delta_1=100$ GeV, left panel) and LH3 ($\delta_1=50$ GeV, right panel). 
The evidence in this figure suggests that in cases in which 
$\delta_1 > M_W$, imposing a minimum 
threshold for $M_{T_W}$ can be helpful in reducing the dominant $WZ/\gamma^\ast$ 
background.  Conversely, when $\delta_1 < M_W$ an upper limit on $M_{T_W}$ can
likewise be of use.  Motivated by these considerations, we allow for either   
a minimum or a maximum threshold for $M_{T_W}$ in our event-selection criteria, 
and only retain events for which  
\begin{itemize}
\item  $M_{T_W} \geq M_{T_W}^{\mathrm{min}}$~~~~ or~~~~ $M_{T_W} \leq M_{T_W}^{\mathrm{max}}$,
\end{itemize}
depending on the benchmark point in question.
As we shall see, such cuts on $M_{\ell_Z\ell_Z}$ and $M_{T_W}$ will turn out to be 
particularly useful in distinguishing a trilepton signal from the dominant 
$WZ/\gamma^\ast$ background.
     
It can also be useful to impose a more stringent lower limit 
$p_{T_{\ell}}^{\mathrm{min}} $ on the transverse momentum $p_{T_{\ell}}$ 
of the charged leptons than that imposed at Level~I:  
\begin{itemize}
\item $p_{T_{\ell}}\geq p_{T_{\ell}}^{\mathrm{min}}>15$~GeV.
\end{itemize}
Likewise, a cut on the total-transverse-momentum variable $H_T$:
\begin{itemize}
\item $H_T \geq H_T^{\mathrm{min}}$,
\end{itemize} 
with $H_T$ defined in terms of the sum
\begin{equation}
  H_T = \displaystyle{\not}E_T + \sum_{i=1}^3 |p_{T\ell_i}|,
\end{equation}
can also be useful in 
differentiating signal 
from background.  A roster of the particular cuts implemented for
each benchmark used in our analysis is compiled in 
Table~\ref{tab:CutParams}. 

\begin{table}
\begin{center}
\begin{tabular}{|c|ccc|c|cc|}\hline
~~Benchmark~~ &
  ~~~$M_{\ell_Z\ell_Z}^{\mathrm{max}}$~~~ &
  ~~~$M_{T_W}^{\mathrm{min}}$~~~ & ~~~$M_{T_W}^{\mathrm{max}}$~~~ & 
  ~~$\Delta R_{\ell\ell}^{\mathrm{max}}$~~ &  
  ~~~~$H_T^{\mathrm{min}}$~~~~ & 
  ~~~$p_{T\ell}^{\mathrm{min}}$~~~ \\\hline
LH1 &~100~GeV~ & ~90~GeV~ &    $-$    & 1.6 & ~240~GeV~ &   $-$    \\
LH2 & ~65~GeV~ &    $-$   &  ~60~GeV~ & 1.3 & ~150~GeV~ &   $-$    \\
LH3 & ~50~GeV~ &    $-$   &  ~60~GeV~ & 1.2 & ~140~GeV~ &   $-$    \\
LH6 & ~65~GeV~ &    $-$   &    $-$    & 1.1 & ~200~GeV~ & ~20~GeV~ \\
LH7 &~100~GeV~ &    $-$   &  ~65~GeV~ & $-$ & ~200~GeV~ &   $-$    \\
LH8 & ~40~GeV~ &    $-$   &    $-$    & $-$ &    $-$    &   $-$    \\
\hline
\end{tabular}
\caption{A list of the optimized Level~III cuts used in the analysis of each 
of the benchmark  points presented in Table~\ref{tab:BMs}. An entry of ``$-$"  
indicates that the corresponding cut is not imposed.  
For more details on the definition of the thresholds used, see text.}
\label{tab:CutParams}
\end{center}
\end{table}

\section{Results\label{sec:Results}}

\begin{table}
\begin{center}
\begin{tabular}{|c|c|cccc|cc|}\hline
 &\multicolumn{5}{c|}{Level~III Cuts}
&& \\ \cline{2-8}     
~Benchmark~ &~$\sigma_{H^\pm A}$ 
          &~$\sigma_{WZ/\gamma^\ast}$ ~ &~$\sigma_{t\bar{t}(j)}$ ~ 
          &~$\sigma_{Wt(j)}$ ~
          &~$\sigma_{\mathrm{BG}}^{\mathrm{comb}}$ ~
          &~~$S/B$~~&~~$S/\sqrt{B}$~~~\\
&(fb)&(fb)&(fb)&(fb)&(fb)&&~~($300~\mbox{~fb}^{-1})$~  \\ \hline
LH1 & 0.038 & 0.159 & 0.020 & 0.011 & 0.191 & 0.20 & 2.15 \\
LH2 & 0.078 & 0.073 & 0.019 & 0.021 & 0.114 & 0.68 & 5.64 \\
LH3 & 0.035 & 0.093 & 0.023 & 0.014 & 0.131 & 0.27 & 2.36 \\
LH6 & 0.101 & 0.185 & 0.030 & 0.007 & 0.221 & 0.46 & 5.27 \\
LH7 & 0.270 & 7.137 & 0.084 & 0.038 & 7.259 & 0.04 & 2.45 \\
LH8 & 0.031 & 0.385 & 0.144 & 0.061 & 0.591 & 0.05 & 1.00 \\ \hline
\end{tabular}
\caption{Cross-sections for the signal process
$pp\rightarrow AH^\pm\rightarrow \ell^+\ell^-\ell^\pm + \displaystyle{\not} E_T$
and for the dominant SM backgrounds from $WZ/\gamma^\ast$, $t\bar{t}(j)$ and $Wt(j)$
production for each of the benchmark points presented in Table~\ref{tab:BMs},
after the application of our Level~III cuts.  The total background 
cross-section is also shown. 
The last two columns display the signal-to-background ratio 
$S/B$, and the statistical significance (as given by $S/\sqrt{B}$)
corresponding to an integrated luminosity of 
${\cal L}=300\mbox{ fb}^{-1}$ in each detector at the LHC 
(operating at a center-of-mass energy
$\sqrt{s}=14$~TeV), after the application of these same cuts.}
\label{tab:BMXSecscutIII}
\end{center}
\end{table}

In Table~\ref{tab:BMXSecscutIII}, we show the discovery potential for the 
trilepton signal at the LHC (assuming a center-of-mass energy of 14~TeV) for 
each of the IDM benchmark points defined above, assuming
an integrated luminosity of 300~${\rm fb}^{-1}$ in each of the two detectors. 
The best prospects for discovery are obtained for the benchmarks LH2 and LH6, each of  
which yields a statistical significance of more than $5\sigma$.
The reason why these benchmarks are comparatively auspicious is twofold.  First, 
both involve a light LIP, with a mass $m_S\sim 40$~GeV.  Second, both also feature a 
mass splitting $\delta_2\sim 70$~GeV, which, on the one hand, is small enough that 
$A\rightarrow SZ\rightarrow S\ell^+\ell^-$ decays will occur through an off-shell 
$Z$ boson, but, on the other hand, is large enough so that the resulting charged
leptons will not generally be too soft to escape detection.

For LH7, which features a similarly light LIP, with $m_S \sim 40$~GeV, but for
which $(\delta_1, \delta_2)=(70,100)$~GeV, the
primary difficulty in resolving the signal is that the (dominant) 
$WZ/\gamma^\ast$ background cannot be suppressed by applying a $Z$ veto on
$M_{\ell_Z\ell_Z}$, since $A\rightarrow SZ\rightarrow S\ell^+\ell^-$ decays occur 
via an on-shell $Z$.  Indeed, this two-body decay mode of the $A$ is analogous to 
what are often referred to as ``spoiler'' processes in the literature on 
trilepton signals in weak-scale supersymmetry~\cite{trilepton}.  Thus,   
although the signal cross section for LH7 after cuts is relatively large, 
the unsuppressed Standard-Model $WZ/\gamma^*$ background renders discovery via 
this channel difficult.  As for LH1, for which $\delta_{1,2}>M_{W,Z}$, the 
Standard-Model $WZ/\gamma^*$ background can be suppressed by imposing a lower limit 
on $M_{T_W}$.  The signal cross section, however, is very small after the imposition 
of this cut, which renders a discovery via this channel difficult for this benchmark
scenario as well.

The discovery prospects
for benchmark point LH3 are also less auspicious.  One reason for this is that
the LIP mass is far heavier in this scenario, and the production cross-section is
therefore appreciably lower, as indicated in Table~\ref{tab:BMXSecs}.  Another is that
since $\delta_1$ and $\delta_2$ are smaller for this benchmark than for LH1 and LH6,
the charged leptons will be significantly softer, and more of them will escape detection.
For this reason, a proportionally greater reduction in signal events occurs as a result of
our detector-acceptance cuts, as can be seen from Table~\ref{tab:LevelIICuts}.  For
benchmark point LH8, $\delta_2$ is smaller still, and the effect of the Level~I cuts 
even more severe; hence the trilepton signal is even more difficult to resolve.

A few further remarks comparing and contrasting the trilepton phenomenology 
of the IDM with that of supersymmetric models are in order.
Indeed, the process 
$pp\rightarrow H^\pm A\rightarrow \ell^+\ell^-\ell^\pm+\displaystyle{\not}E_T$, 
which yields the dominant contribution to the trilepton signal in the IDM 
is in many ways analogous to the direct chargino-neutralino 
production process $pp \rightarrow \chi_2^0 \chi_1^\pm$, 
with $\chi_2^0\rightarrow \chi_1^0 Z^{(*)} \rightarrow 
\chi_1^0 \ell^+ \ell^-$ and $\chi_1^\pm\rightarrow \chi_1^0 W^{(\pm*)} \rightarrow 
\chi_1^0 \ell \nu$, 
where $\chi_{1,2}^0$ are the lightest and second lightest neutralinos and 
$\chi_1^\pm$ is the lightest chargino.
This channel has long been regarded as a promising discovery channel for
weak-scale supersymmetry.  Indeed, as was shown in~\cite{XerxesLHCTrilepton}, for 
certain opportune regions of parameter space, an observable signal could be obtained
with less than $30\mbox{~fb}^{-1}$ of integrated luminosity at the LHC. 
More recently, the CMS collaboration, working in the context of minimal supergravity, 
has indicated that a $5\sigma$ discovery of supersymmetry could be achieved in this channel
with $30\mathrm{~fb}^{-1}$ of integrated luminosity, provided that the gaugino
mass parameter $M_{1/2}\lesssim 180$~GeV~\cite{CMSTDR}.    
 
Thus, we see that given similar mass spectra, the discovery prospects 
for the supersymmetric process are markedly better than those for its IDM counterpart.  
This is primarily due to  
to the substantial difference --- a relative factor of around 16 --- between the 
production cross-sections for $pp \rightarrow \chi_2^0 \chi_1^\pm$ in the 
minimal supersymmetric Standard Model (MSSM) and $pp\rightarrow AH^\pm$ in the 
IDM.  This difference owes to two important distinctions between the 
characteristics of the relevant particles in the
two models.  The first of these is that  
$\chi_1^\pm$ and $\chi_2^0$ are Weyl fermions whereas $H^\pm$ and $A$ are 
real scalars.  As a consequence, the cross-sections for the corresponding 
processes in the two models differ by a relative factor of roughly 4 in the 
high-energy limit (i.e.\ the limit in which $s \gg m_{i}^2$, where $m_i$ 
denotes the mass of any of the particles involved in the interaction). 
The second relevant distinction is that 
the scalar doublet $\phi_2$ of the IDM is in the fundamental representation 
of $SU(2)$, whereas the charged and neutral Winos (which respectively constitute the 
dominant components of $\chi_1^\pm$ and $\chi_2^0$ in the relevant region of SUSY 
parameter space) are in the adjoint representation.  This translates into another 
relative factor of 4 between the corresponding production cross-sections.
The practical consequence of this result, of course, is that observing a trilepton
signal in the IDM is far more difficult than it is in its MSSM analogue.
Indeed, we have seen that
although the trilepton channel is one of the cleanest channels in which one might hope
to discover supersymmetry at the LHC, in the IDM, this channel can only be observed 
in the region of parameter space in which the LIP is light ($m_S\sim 40$ GeV) and the 
mass splitting $\delta_2$ is relatively large ($\delta_2 \sim 70$ GeV).
 
While the above analysis was performed assuming a center-of-mass energy 
$\sqrt{s}=14$~TeV, it is also worthwhile to consider how the discovery prospects
would differ at an LHC operating energy of $\sqrt{s}=10$~TeV.  
In this case, the $pp\rightarrow H^\pm A$ production cross-sections 
are reduced to roughly 60\% of the values given in Table~\ref{tab:BMXSecs}, 
while the (generally dominant) $WZ/\gamma^\ast$ background drops to roughly 80\%
of its 14~TeV value.  Since signal event count is not generally a limiting factor in
event selection, we would expect each of the $S/\sqrt{B}$ values quoted above 
to drop to roughly 65\% of its 14~TeV value at a 10~TeV machine, given identical
luminosities and assuming similar cut efficiencies.  
While this is not an imperceptible reduction, it is by no means a
severe one; thus, were our universe in fact to resemble that described by an IDM
benchmark scenario such as LH2 or LH6, one would still expect to see evidence 
of trilepton production from the decays of heavy inert particles at the LHC, even
at $\sqrt{s}=10$~TeV.

\section{Conclusions\label{sec:Conclusion}}

The Inert Doublet Model is a simple yet incredibly versatile scenario for 
physics beyond the Standard Model.  Among its phenomenological advantages is that  
it provides a viable WIMP dark-matter candidate in the form of the lightest inert 
particle.  
In this work, we have investigated the observability of a trilepton signal 
at the LHC in the Inert Doublet Model.
While the first signals of an inert doublet at the LHC are likely to appear in the
dilepton channel \cite{ArizonansDilepton}, the observation of a signal in 
the trilepton channel could 
provide valuable additional information about the parameter space of the model   
and assist in distinguishing the IDM from other BSM 
scenarios which give rise to
similar signature patterns.  We have shown that at an integrated luminosity
$\mathcal{L}=300\mbox{~fb}^{-1}$, it should be possible to resolve the trilepton 
signal, provided that the LIP is light ($m_S \sim 40$~GeV), the mass splitting
$\delta_2$ lies within the range $50\mbox{~GeV}\lesssim\delta_2\lesssim M_Z$, 
and $\delta_1$ is small enough ($\delta_1\lesssim 100$~GeV) that $H^\pm A$ 
production is not drastically suppressed.  These criteria coincide with those 
which lead to the best detection prospects in the dilepton channel as well.

It should be noted, however, that one could only hope to observe a trilepton signal 
in regions of parameter space in which the Higgs is lighter than around 180~GeV.  
Although the Inert Doublet Model can certainly accommodate a heavier Higgs boson ---
indeed, among the model's numerous advantages is its ability to alleviate the 
little hierarchy problem --- the requisite contributions to the oblique $T$ parameter
needed for this owe to the existence of a sizeable 
mass splitting between $H^\pm$ and $S$.
When this is the case, the $pp\rightarrow AH^\pm$ production cross-section will 
be highly suppressed, and the trilepton-signal contribution from this process 
will consequently be unobservable at the LHC.  

As a final word, we note that although the analysis performed in this 
work was conducted within the framework 
of the Inert Doublet Model, similar signatures involving the production of charged 
and neutral scalars which subsequently decay into other, lighter, scalar particles 
and SM gauge bosons appear in many other BSM scenarios.  We
emphasize that our results should also apply in any such scenario in which the 
aforementioned lighter scalar particle is neutral and stable (on collider time scales), 
and hence appears in the detector as missing energy.  The observation of a clean trilepton$~+~\displaystyle{\not}E_T$ signal above the SM background would be a clear 
indication of new physics.   To determine the precise nature of that 
new physics, however, and to pin down the particle nature 
of the dark matter candidate will likely require additional data from a variety of
sources.  These may include complementary channels at the LHC, signals at 
direct or indirect dark-matter-detection experiments, or results from one of many 
other available experimental probes of physics beyond the Standard Model.


\section{Acknowledgments}
We would like to thank J. Alwall for correspondence regarding the MadGraph simulation
package, and E. Dolle for useful discussions.  This work was supported 
in part by the Department of Energy under Grant~DE-FG02-04ER-41298.


\smallskip


\end{document}